\documentclass[modern]{aastex631}
\usepackage{amsmath}

\shorttitle{Long Stellar Rotation Periods from TESS}
\shortauthors{Hattori et al.}

\begin{document}

\title{Measuring Long Stellar Rotation Periods ($>$10 days) from TESS FFI Light Curves is Possible: An Investigation Using TESS and ZTF}

\correspondingauthor{Soichiro Hattori}
\email{soichiro.hattori@columbia.edu}

\author[0000-0002-0842-863X]{Soichiro Hattori}
\affiliation{Department of Astronomy, Columbia University, 538 West 120th Street, Pupin Hall, New York, NY 10027, USA}
\affiliation{Department of Astrophysics, American Museum of Natural History, 200 Central Park West, New York, NY 10024, USA}

\author[0000-0003-4540-5661]{Ruth Angus}
\affiliation{Department of Astrophysics, American Museum of Natural History, 200 Central Park West, New York, NY 10024, USA}

\author[0000-0002-9328-5652]{Daniel Foreman-Mackey}
\affiliation{Center for Computational Astrophysics, Flatiron Institute, 162 5th Avenue, New York, NY 10010, USA}

\author[0000-0003-4769-3273]{Yuxi (Lucy) Lu}
\affiliation{Department of Astronomy, The Ohio State University, Columbus, 140 W 18th Ave, OH 43210, USA}
\affiliation{Center for Cosmology and Astroparticle Physics (CCAPP), The Ohio State University, 191 W. Woodruff Ave., Columbus, OH 43210, USA}
\affiliation{Department of Astrophysics, American Museum of Natural History, 200 Central Park West, New York, NY 10024, USA}

\author[0000-0001-8196-516X]{Isabel Colman}
\affiliation{Liberal Studies, New York University, 726 Broadway, New York, NY 10003, USA}
\affiliation{Department of Astrophysics, American Museum of Natural History, 200 Central Park West, New York, NY 10024, USA}



\begin{abstract} 
The rotation period of a star is an important quantity that provides insight into its structure and state. For stars with surface features like starspots, their periods can be inferred from brightness variations as these features move across the stellar surface. TESS, with its all-sky coverage, is providing the largest sample of stars for obtaining rotation periods. However, most of the periods have been limited to shorter than the 13.7-day TESS orbital period due to strong background signals (e.g., scattered light) on those timescales. 
In this study, we investigated the viability of measuring longer periods ($>10$ days) from TESS light curves for stars in the Northern Continuous Viewing Zone (NCVZ). We first created a reference set of 272 period measurements longer than 10 days for K \& M dwarfs in the NCVZ using data from the Zwicky Transient Facility (ZTF) that we consider as the ``ground truth" given ZTF's long temporal baseline of 6+ years. We then used the \texttt{unpopular} pipeline to de-trend TESS light curves and implemented a modified Lomb-Scargle (LS) periodogram that accounts for flux offsets between observing sectors. For 179 out of the 272 sources (66\%), the TESS-derived periods match the ZTF-derived periods to within 10\%. The match rate increases to 81\% (137 out of 170) when restricting to sources with a TESS LS power that exceeds a threshold. Our results confirm the capability of measuring periods longer than 10 days from TESS data, highlighting the dataset's potential for studying slow rotators.
\end{abstract}


\section{Introduction} \label{sec:intro}
As the loss of angular momentum is a key process in the life of cool main-sequence stars (i.e., G, K, M dwarfs), their rotation periods provide information about the stellar structure and evolutionary state. The primary mechanism for stellar spin down is \textit{magnetic/rotational braking}, where the magnetic field produced by the convective envelope of these low-mass stars interact with the stellar wind to transport angular momentum away from the star \citep{schatzman1962, brandt1966, webdav1967}.

An important property that can be inferred from rotation periods is the star's age.
While contemporaneously born stars in open clusters can initially have a spread in their rotation periods, such as the bimodal distribution of rotation periods observed among T Tauri stars in the Orion Nebula cluster \citep{Attridge1992, choi1996}, the rotation periods of solar-type stars are known to converge after ${\sim}1$ Gyr (i.e., they lose information about their initial rotation period) \citep{endal1981, stauffer1987, keppens1995}. This convergence, combined with the rotation-age relationship $\Omega \propto t^{-1/2}$, where $\Omega$ is the angular velocity of the star and $t$ is its age, implied by the magnetized stellar wind model of \citet{webdav1967} and empirically supported by \citet{skumanich1972}, gave rise to \textit{gyrochronology} \citep{barnes2003}, the method of age dating low-mass stars from their rotation periods. While not all the physics of this process are understood with recent observations providing evidence for deviations from the rotation-age relationship given above, gyrochronology is still considered one of the more precise methods for inferring the ages of individual solar-type stars \citep{soderblom2010}.

Stellar rotation is also related to the strength of a star's magnetic field and consequently its \textit{activity} (e.g., starspots, faculae). Specifically, there exists an inverse correlation between a star's fractional X-ray luminosity $L_\mathrm{X} / L_\mathrm{bol}$, where $L_\mathrm{X}$ is the star's X-ray luminosity and $L_\mathrm{bol}$ is its bolometric luminosity, and its \textit{Rossby number} $R_\mathrm{0} = P_\mathrm{rot} / \tau$, the ratio between its rotation period $P_\mathrm{rot}$ and its mass-dependent convective turnover time $\tau$ \citep{pallavicini1981, noyes1984, patten1996}. While this relationship is known to break down for fast rotators \citep{micela1985}, where the fractional X-ray luminosity saturates at ${\sim}10^{-3}$ \citep{vilhu1984}, for slow rotators it allows us to infer the magnetic dynamo processes \citep{wright2011}. 

Stellar ages and activity indicators are important quantities that are useful beyond the field of stellar astrophysics. Stellar ages are crucial quantities in galactic dynamics and galactic archaeology, the study of the history and evolution of the Milky Way, as it would allow for confirming whether predictions of the past movement of stellar populations match our observations \citep{edvardsson1993, freeman2002, casagrande2011, jofre2021, lu2024}. In the field of exoplanets, stellar ages would allow for estimating the age of observed exoplanets, and activity indicators would allow for a more holistic inference on their atmospheric properties and potential habitability beyond simply orbiting in the annulus where liquid water could exist \citep{rushby2013, gaidos2023}. Hence, stellar rotation period measurements are beneficial for a wide range of scientific investigations.



With many large (photometric) sky surveys occurring in the past few decades, including CoRot \citep{baglin2006}, Kepler \& K2 \citep{borucki2010, howell2014}, the Transiting Exoplanet Survey Satellite (TESS) \citep{ricker2015}, the Zwicky Transient Facility (ZTF) \citep{bellm2019}, and the Gaia satellite \citep{gaia2016}, period measurements from brightness variations caused by the stellar surface inhomogeneities (e.g., starspots, faculae) coming and going out of view \citep{kron1947} has been a fruitful endeavor. Systematic searches in these datasets have produced several large rotation period catalogs including (but certainly not limited to) those by \citet{mcquillan2014, chen2020, santos2019, santos2021, lu2022, holcomb2022, gaia2023, colman2024}.


We follow this line of work of detecting periodic photometric modulations and report rotation periods from TESS. TESS \citep{ricker2015}, with its precise pointing, high cadence, and all-sky coverage, has already provided the opportunity to measure rotation periods from the largest sample of stars to date (e.g., \citealt{healy2020, canto2020, kounkel2022, holcomb2022, fetherolf2023, colman2024, claytor2022, claytor2024, binks2024}). However, due to the presence of systematic effects such as the scattered light\footnote{\url{https://tess.mit.edu/observations/scattered-light/}} from the Earth and Moon increasing the background flux levels roughly every 13.7 days (i.e., the orbital period of the satellite), the lack of measurements during the data downlink between every orbit, and its 27-day observational baseline (i.e., TESS sector) for a single pointing, besides \citet{claytor2022, claytor2024, colman2024}, period measurements from TESS have largely been limited to $\lesssim$13 days. This limitation to shorter periods is unfortunate as even though TESS has two continuous viewing zones (CVZs), $24^\circ$-diameter circular regions around the north and south ecliptic poles that are continuously observed by TESS (excluding the data-downlink gaps) for ${\sim}1$ year at a time, we are not taking full advantage of them to study the slow rotators which are generally underrepresented due to the need for observations over a long temporal baseline. 

Therefore, in this study, we investigate the viability of measuring \textit{long rotation periods} (i.e., $>$10 days) with TESS. We note that while our study is similar in spirit to \citet{claytor2024} as we are also measuring long rotation periods from stars in the CVZs, our work is complementary as the target sample and methodology are different. \citet{claytor2024} focuses on sources in the southern CVZ while we focus on those in the northern CVZ. Also, \citet{claytor2024} detects periods by using convolutional neural networks (CNNs) \citep{lecun1989}, a machine learning algorithm which has been successfully used in a wide range of problems both in academia and beyond. As CNNs generally take 2D image data as input, \citet{claytor2024} performed wavelet transforms on the light curves to generate 2D images (frequency vs time) and performed their inference in the transformed space. Compared to working in this transformed 2D image space, our approach is similar to previous rotation period studies as we work in the flux vs. time space.

 As opposed to a systematic search to detect potential new slow rotators and produce a catalog, the objective of this study is to \textit{establish that long rotation periods can be measured from the TESS prime mission full-frame image (FFI) light curves in the northern CVZ (NCVZ)}. This objective therefore requires a sample of stars that have known long rotation periods. To select this ``ground-truth'' sample, we use the publicly available ZTF data to detect slow rotators using Lomb-Scargle (LS) periodograms \citep{lomb1976, scargle1982, vanderplas2018}. ZTF \citep{bellm2019}, a ground-based optical telescope located at the Palomar Observatory, has been conducting a survey since March 2018 which surveys the entire northern sky every two days. The roughly daily observations, 6+ year long temporal baseline, and the contemporaneous observations of the northern sky with TESS \citep{roestel2019, prince2020} make ZTF an excellent resource to select our sample from. After using ZTF data to detect slow rotators in the TESS NCVZ, we attempt to recover the same period from the TESS Cycle 2 FFI data. We use the \texttt{unpopular} \citep{hattori2022} package, a Python library for removing systematics from FFI light curves by performing $L_2$-regularized linear regression. Each of these steps will be discussed in subsequent sections. Compared to previous de-trending pipelines that either explicitly (e.g., by fitting a low-order polynomial to the light curve) or implicitly removed stellar variability as they were designed for detecting and characterizing transit signals, \texttt{unpopular} attempts to retain these signals. 

We show that it is possible to measure long rotation periods from TESS FFI light curves, even exceeding the length of a single TESS sector (i.e., 27 days). Of the sample of 272 potential slow rotators selected according to the ZTF $r$-band data, we achieve a 66\% (179/272) match rate between the ZTF and TESS periods. Roughly half (43 sources) of the 93 non-matches were cases where the TESS-measured period was half the ZTF-measured period, a common failure mode for LS periodograms \citep{vanderplas2018}.
If we only keep sources with a with maximum TESS Lomb-Scargle power greater than 0.02, we achieve a 81\% (137/170) match rate. 
We note that our sample size is relatively small with $N=272$ sources as the objective of this study is not to produce a catalog of newly measured rotation periods but to instead show that reliable long periods can indeed be measured by TESS. A catalog will be produced in an upcoming paper.

This paper is organized as follows. In Section \ref{sec:ztf} we discuss the ZTF data, the periods measured from it, and the final target list that is then analyzed with TESS. Section \ref{sec:tess} discusses the TESS data, the \texttt{unpopular} pipeline, and our modification to the LS periodogram to handle flux offsets between TESS sectors. We compare our measured periods between ZTF \& TESS, report our recovery rate, and discuss limitations of our study in Section \ref{sec:results}. We conclude and explore future work in Section \ref{sec:conclusion}. 

\section{ZTF Data \& Periods} \label{sec:ztf}
\subsection{Zwicky Transient Facility} \label{sec:ztf_info}

The Zwicky Transient Facility (ZTF) \citep{bellm2019} is a ground-based 48-inch optical telescope located at the Palomar Observatory. With its extremely wide 47 square degree field-of-view (FOV), ZTF is able to survey the entire northern sky every two days in ZTF-\textit{r} (${\sim}410\text{-}550\,\mathrm{nm}$) and ZTF-\textit{g} (${\sim}560\text{-}730\,\mathrm{nm}$)
with a saturation \& limiting magnitude for both bands of ${\sim}$13 \& ${\sim}$20, respectively \citep{dekany2020}. While ZTF also observes in the \textit{i}-band (${\sim}720\text{-}900\,\mathrm{nm}$), only the $r$ \& $g$ band measurements are publicly released. As ZTF had started observing in March 2018, it has already observed and released measurements spanning over 6 years of the northern night sky. Given this long temporal baseline, we use ZTF data to initially detect slow rotators within the sample we describe below and treat these measured periods as the ``ground-truth" periods we would like to recover from TESS FFI light curves.

\subsection{Target List Creation} \label{sec:target_list}
As the sources for this study must have both ZTF observations and multiple sectors of TESS observations, we started by first selecting a sample of stars in the TESS northern continuous viewing zone (NCVZ), a $24^\circ$-diameter circular region centered at the northern ecliptic pole (RA, Dec = $270.0^\circ$, $+66^\circ\,33'\,39''$) where TESS continuously observed for ${\sim}$1 year during its second year. We restricted the sample to K \& M dwarfs as those have a higher chance of exhibiting slow brightness variations (i.e., long periods)  \citep{giles2017}. Using \texttt{Astroquery} \citep{astroquery2019} we selected sources in the NCVZ from the TESS \textbf{Candidate Target List (CTL) v8.01} \citep{stassun2018, stassun2019} satisfying the following criteria: 
\begin{itemize}
    \vspace{-0.2cm}\item $ 2000\,\mathrm{K} < T_\mathrm{eff} < 5300\,\mathrm{K}$
    \vspace{-0.2cm}\item $13 < G < 16$
    \vspace{-0.2cm}\item \texttt{objType = STAR}
    \vspace{-0.2cm}\item \texttt{lumclass = DWARF}
    \vspace{-0.2cm}\item \texttt{contratio}\footnote{The contamination ratio (\texttt{contratio}) is defined as the ratio of the flux of contaminants to the flux of the source of interest. Contaminants are searched for within 10 TESS pixels of the target. More information is provided in Section 3.3.3 of \citet{stassun2018}.} $< 0.05$
\end{itemize}
where $G$ is the Gaia G magnitude \citep{gaia2016}. We opted to use the CTL, a substantially smaller subset of the TESS Input Catalog (TIC), as the CTL calculates the contamination ratio and also performs cuts on the TIC to increase the chance that an observed source is a main-sequence star (e.g., not a white dwarf). While the CTL has a magnitude cut removing stars fainter than 13th magnitude, possibly removing M dwarfs, it overrides this cut by explicitly including sources in the specifically curated Cool Dwarf list \citep{muirhead2018}.

This initial query returned 9612 sources. We then obtained Gaia information of these sources by querying the Gaia Data Release 3 (DR3) source table \citep{gaiadr3_2023} and removed sources with Gaia Renormalized Unit Weight Errors (RUWE) larger than 1.4 as higher values indicate potential binaries. This RUWE cut left us with 7804 sources.

\subsection{Downloading ZTF Data} \label{sec:ztf_download}
We then downloaded the $r$-band and $g$-band light curves from the ZTF Public Data Release 22 dataset using the ZTF-LC-API\footnote{\url{https://irsa.ipac.caltech.edu/docs/program_interface/ztf_lightcurve_api.html}} and querying for sources within a circle of radius $1.8\arcsec$ of their Gaia DR3 positions. ZTF DR22 contains measurements from March 17th, 2018 (MJD $\geq 58194.0$) until June 30th, 2024 (MJD $\leq 60491.0$). We follow the data release notes\footnote{Section 12 of \url{https://irsa.ipac.caltech.edu/data/ZTF/docs/releases/dr22/ztf_release_notes_dr22.pdf}} and set \texttt{BAD\_CATFLAGS\_MASK=32768} to mask bad/unusable data. We also set \texttt{NOBS\_MIN=1000} to require each source to have at least 1000 observations. However, as this number of observations is of \textit{all} measurements, including those from private surveys, we removed sources with less than 500 public $r$-band measurements to lower spurious period detections. We made cuts based on the $r$-band measurements as they are taken at a higher cadence and are less susceptible to observing conditions compared to the $g$-band\footnote{\url{http://nesssi.cacr.caltech.edu/ZTF/Web/gettingto1.html\#calib}}.  These cuts reduced the sample size to 6537 sources.

\subsection{ZTF-measured Periods} \label{sec:ztf_periods}
\begin{figure}
    \centering
    \includegraphics[width=\textwidth]{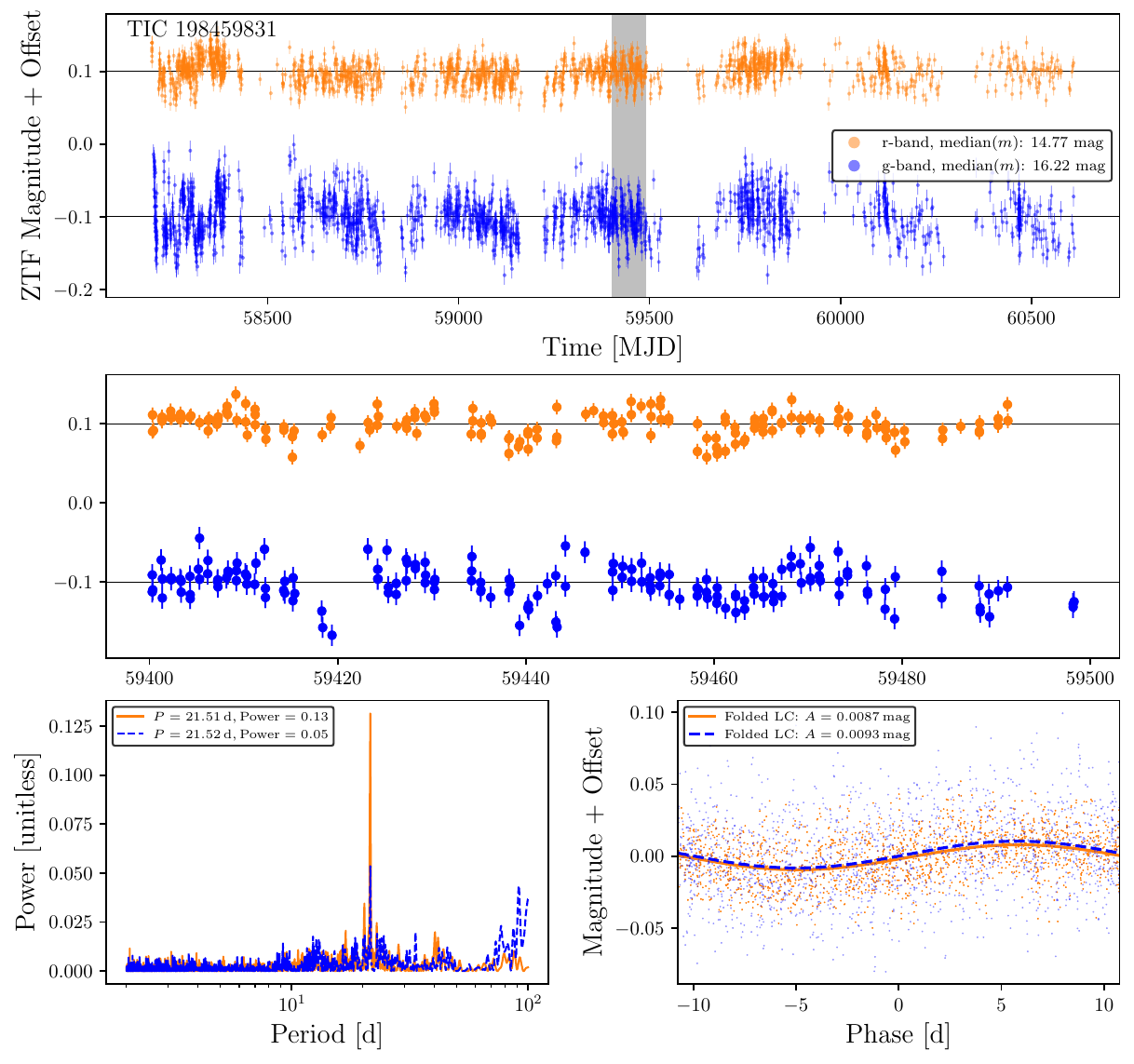}
    \caption{The ZTF $r$ \& $g$ band light curves and the period analysis for TIC 198459831. In the top panel we show the $r$ and $g$ band light curves. The middle panel shows the same light curves but zooms into the shaded region between 59400-59500 days.
    The bottom left panel shows the LS periodogram over the range of 2-100 days. In the bottom right panel we show the phase-folded light curve, folded at the period with the highest LS power, and overlay the sinusoidal model at this period. We also indicate $A$, the semi-amplitude of the oscillation.}
    \label{fig:ztf_summary_plot}
\end{figure}

\begin{figure}
    \centering
    \includegraphics[width=\textwidth]{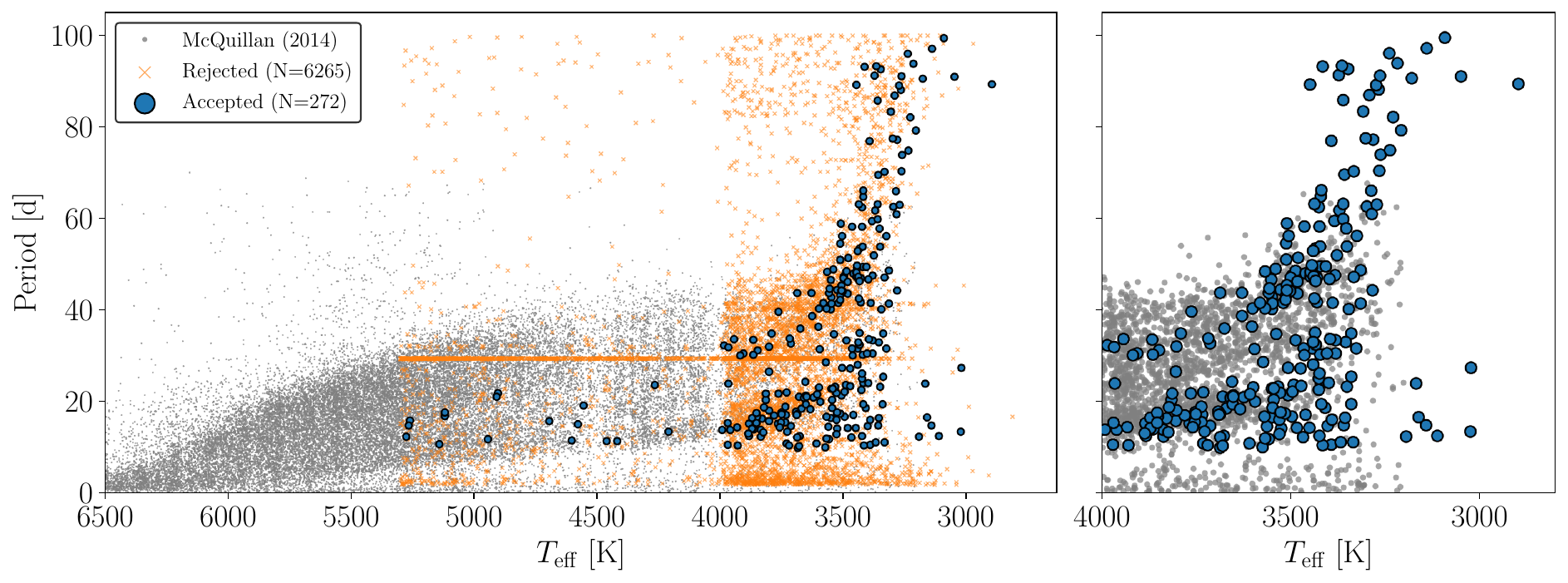}
    \caption{The measured ZTF periods overlaid on the \citet{mcquillan2014} Kepler catalog. In the left panel, the orange crosses ($N=6265$) indicate targets which did not pass the cut described in the text, while the blue circles ($N=272$) indicate those that did. The horizontal streak of periods at ${\sim}29$ days is due to the lunar phase and are treated as systematics. The high temperature cutoff is due to our threshold of $T_\mathrm{eff} < 5300\,\mathrm{K}$ and the increase of sources at $T_\mathrm{eff} < 4000\,\mathrm{K}$ is due to the way the TESS CTL is produced where the Cool Dwarf list is included \citep{muirhead2018}. The right panel is the same as the left panel but zoomed into the region where $T_\mathrm{eff} < 4000\,\mathrm{K}$. We can see that our accepted periods trace out the shape present in \citet{mcquillan2014}. }
    \label{fig:ztf_over_mcquillan}
\end{figure}

Before the period search, we preprocessed the ZTF light curves by sigma clipping at the 3-$\sigma$ level to remove outliers and subtracted a linear trend from the data. We then used the \texttt{Astropy} \citep{astropy:2013, astropy:2018, astropy:2022} implementation of the Lomb-Scargle (LS) periodogram to detect sinusoidal signals. We searched for frequencies corresponding to periods between 2-100 days (1-day signals are systematics due to the ZTF observing cadence) in the $r$-band light curves. While we only care about sources with $P_\mathrm{rot} > 10\,\mathrm{day}$, we included shorter periods to  reduce the chance of having short-period sources in the sample. We also set the \texttt{samples\_per\_peak=50} argument to ensure that the frequency/period around each peak is sampled finely. The frequency/period with the highest LS power was considered the star's measured period. As we are treating these ZTF measured periods as the ground truth, we only retained the sources which satisfied the following (restrictive) criteria in the $r$-band light curve:  
\begin{itemize}
    \vspace{-0.2cm}\item $P_\mathrm{rot} > 10\,\mathrm{day}$
    \vspace{-0.2cm}\item \texttt{max\_power} $> 0.1$
    \vspace{-0.2cm}\item \texttt{SNR} $> 50$
\end{itemize}
where \texttt{max\_power} is the maximum peak power returned by the LS periodogram and \texttt{SNR} (signal to noise ratio) is the ratio of the max power to the median power of the periodogram. Of the 380 sources that passed the above cut, we also removed 108 sources with periods that fall within $29\,\mathrm{day}< P_\mathrm{rot} < 30\,\mathrm{day}$ as ZTF light curves are known to show variations due to the lunar phase\footnote{\textbf{Importance of Lunar Phase} section in \url{http://nesssi.cacr.caltech.edu/ZTF/Web/ZuberPars.html}}. These fairly restrictive cuts leaves us with a sample of 272 sources. For completeness we also preprocessed and searched for periods in the $g$-band light curves of these 272 sources. For 240 of the sources we recovered the same period in the $g$-band as the $r$-band to within 10\%. In Figure \ref{fig:ztf_summary_plot} we show an example of a source TIC 198459831 that passed these cuts and measured the same period in the two bands. Of the 32 mismatches, 5 were due to there being no corresponding downloadable $g$-band light curves. Due the aforementioned condition of $g$-band measurements having lower precision than $r$-band measurements we do not have a cut that requires that periods measured from the two bands to be the same. As a sanity check, we overlay these periods on the \citet{mcquillan2014} Kepler catalog to check for bulk consistency (see Figure \ref{fig:ztf_over_mcquillan}). We can see that our accepted periods trace the general shape of the \citet{mcquillan2014} catalog. Table \ref{tab:table_1} contains information on these 272 sources.

\begin{table}[]
\begin{centering}
\caption{Description of columns for the 272 sources used in this work (see Section \ref{sec:target_list})}
\begin{tabular}{lllll}
Column             & Unit & Description &  &  \\
\hline \hline 
\texttt{TIC}               &   \ldots   & TESS Input Catalog ID  &  &  \\
\texttt{GAIA}           &   \ldots   &       Gaia Designation      &  &  \\
\texttt{ra\_dr2}            & deg  &  Gaia DR2 Right Ascension (used in TESS Input Catalog)           &  &  \\
\texttt{dec\_dr2}           & deg  &  Gaia DR2 Declination (used in TESS Input Catalog)           &  &  \\
\texttt{ra\_dr3}            & deg  &  Gaia DR3 Right Ascension           &  &  \\
\texttt{dec\_dr3}           & deg  &  Gaia DR2 Declination            &  &  \\
\texttt{Tmag}               & mag  &  TESS magnitude           &  &  \\
\texttt{e\_Tmag}               & mag  &  TESS magnitude error           &  &  \\
\texttt{Gaia\_g\_mean\_mag} & mag  &  Gaia G-band mean magnitude           &  &  \\
\texttt{Teff}               & K    &  Effective temperature from TESS Input Catalog          &  &  \\
\texttt{e\_Teff}               & K    &  Effective temperature error          &  &  \\
\texttt{Prot\_ZTF\_r}       & d    &  ZTF r-band rotation period           &  &  \\
\texttt{Prot\_TESS}         & d    &  TESS rotation period           &  & \\
\hline \hline \\ 
\end{tabular}
\tablecomments{This table is available in its entirety in machine-readable form.}
\label{tab:table_1}
\end{centering}
\end{table}

\section{TESS Data \& Periods} \label{sec:tess}
For each of the 272 sources with a measured ZTF $r$-band light curve rotation period longer than 10 days (see Section \ref{sec:ztf}), we downloaded the TESS Cycle 2 FFI data for the same source and attempted to measure a rotation period from the de-trended FFI light curve using a modified LS periodogram to handle flux offsets between sectors.

\subsection{TESS \& Downloading FFI Cutouts} \label{sec:tess_data}
TESS \citep{ricker2015} is an all-sky survey with a $24^\circ \times 96^\circ$ FOV, an angular resolution of $21\arcsec$, and a bandpass spanning $600\text{-}1000\,\mathrm{nm}$ that continuously observes a strip of the night sky for ${\sim}27$ days (i.e., TESS observing sector) at a time. The spacecraft is rotated eastward every sector to cover a hemisphere over 13 sectors. During the 2-year primary mission, the southern hemisphere was observed from sectors 1-13 (i.e., Cycle 1) while the northern hemisphere was covered from sectors 14-26 (i.e., Cycle 2) with FFIs at 30-minute cadence. For any given cycle one of the cameras is always observing the ecliptic pole, creating the ${\sim}1$-year continuous viewing zones (CVZ) from where we selected our targets. 

We used the \texttt{lightkurve} package \citep{barentsen2020}, which uses \texttt{TESScut} \citep{brasseur2019}, to download FFI cutouts from each sector (${\sim}27$ days) based on the Gaia DR2 \citep{gaiadr2_2018} positions as the TESS CTL is created with DR2 positions. We checked the Gaia designations to ensure the correct source positions between DR2 \& DR3 were used. As these sources are in the NCVZ (sectors 14-26), most of them have 12-13 sectors worth of data. For each sector of each target, we downloaded $41 \times 41$ pixel FFI cutouts where the target is centered in the cutout. While this size is smaller than the $100 \times 100$ cutouts mentioned in the \texttt{unpopular} paper \citep{hattori2022}, we found that the period measurements obtained from using 128 regressors on a $41 \times 41$ FFI cutout and 256 regressors on a $91 \times 91$ FFI cutout had minimal difference. As such, we opted to use the smaller FFI cutouts for storage convenience. Before de-trending, we removed sectors when the source fell on non-science pixels close to the detector edge or if more than 10\% of flux measurements in a given sector were \texttt{NaN}s due to high levels of saturation. 

\subsection{De-trending} \label{sec:unpop_detrending}
Each sector of data is de-trended independently using the \texttt{unpopular} pipeline. In contrast to approaches such as the Systematics Insensitive Periodogram \citep{angus2016, hedges2020} which can simultaneously de-trend multiple sectors, \texttt{unpopular} only de-trends one sector at a time. While the details of the method are presented in \citet{hattori2022}, the \texttt{unpopular} method, which builds upon the \textsl{Causal Pixel Model} method \citep{wang2016}, creates an $L_2$-regularized linear model to capture the systematics in a given pixel by using other \textsl{astronomically distant} pixels as features. The underlying idea is that since systematic effects (e.g., scattered light) are likely to affect the majority of pixels in the FOV, trends that are common (i.e., popular) across the FFI are systematics and not astrophysical variations. Under this assumption, we can use flux variations from other pixels as features for our model as they will contain the same systematics but not the same astrophysical variations as the pixel of interest. The set of chosen features, in this case 128 other pixels, are chosen based on having a similar median brightness to the pixel being de-trended from the entire FFI cutout \textsl{excluding} a $5 \times 5$ region of pixels surrounding the pixel being de-trended. This exclusion region is necessary to prevent any pixels capturing astrophysical flux variations from the source from being used as features in the systematics model. 

$L_2$ regularization (i.e., ridge regression) \citep{ridge}, which involves adding the square of the sum of the coefficients (i.e., the square of the Euclidean norm of the weight vector $\boldsymbol{w}$) to the objective function, has the effect of shrinking the coefficients toward zero and preventing overfitting. For each source, we use a $3 \times 3$ aperture and de-trend 9 pixels before summing the fluxes together. While it is possible to simultaneously fit a polynomial model as in \citet{hattori2022} in addition to the systematics model described above, for this study we only used the systematics model with a regularization value of $\lambda=0.01$. We settled on this value after de-trending several light curves with eight $\lambda$ values each separated by an order of magnitude starting at $\lambda = 10^{-5}$ and ending at $\lambda = 10^2$, and confirming that the results are similar over a range of $\lambda$ values . Only when $\lambda$ is large (e.g., $10, 100$), which is when the model is strongly regularized, does the light curve look appreciably different. 

In addition to $\lambda$, another hyperparameter for this model is $k$, the number of contiguous sections to divide a given sector's data into. As discussed in section 2.3 of \citet{hattori2022}, this splitting is necessary to ensure that the data used to obtain the model's coefficients (i.e., the training set) and the data being de-trended (i.e., the test set) are mutually exclusive. Experiments related to changing the value of $\lambda$ and $k$ are already presented in \citet{hattori2022}, and an exhaustive search over possible values for finding optimal values are beyond the scope of this study. 

\begin{figure}
    \centering
    \includegraphics[width=\textwidth]{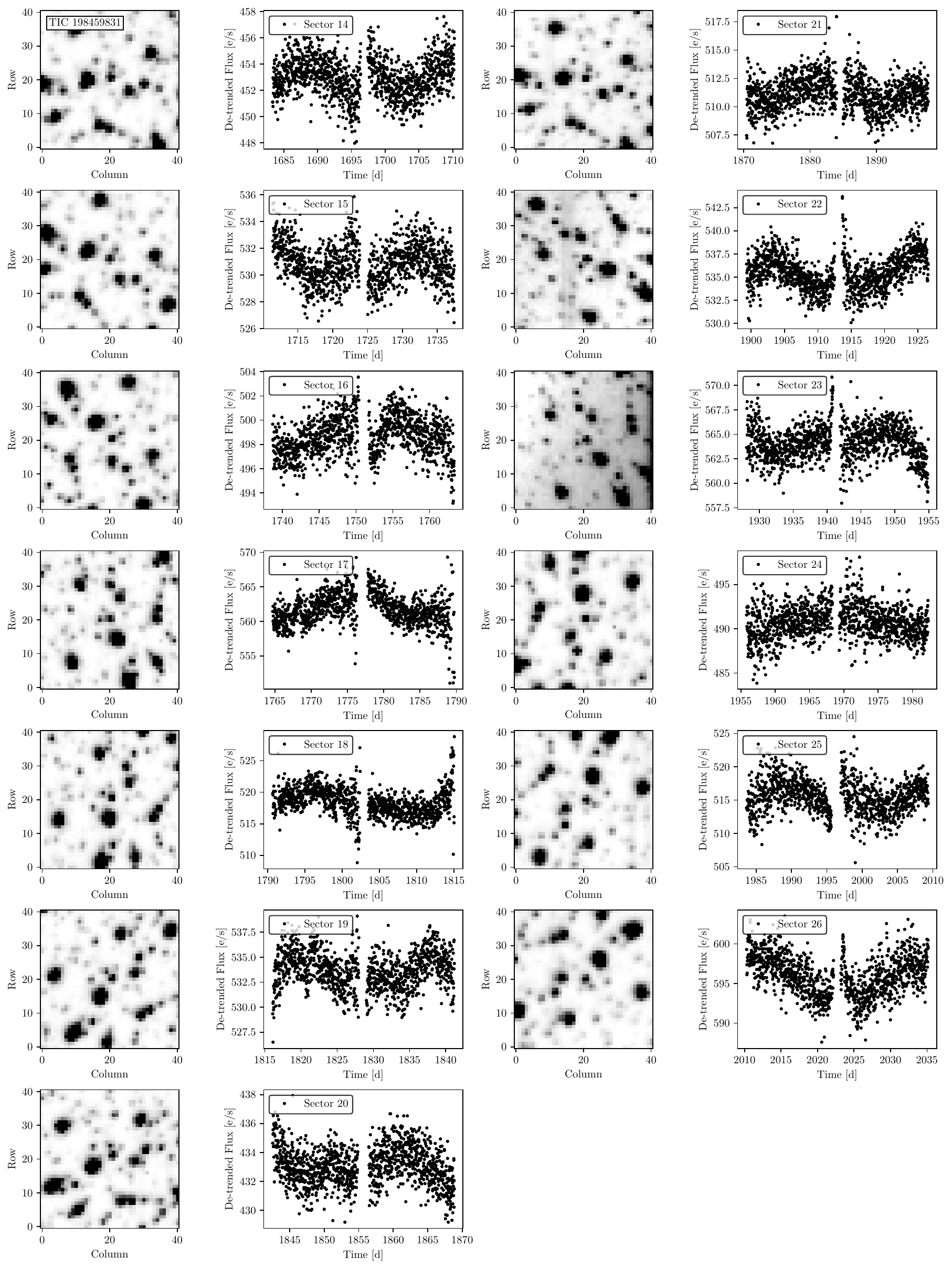}
    \caption{TESS FFI cutouts and de-trended light curves for TIC 198459831. Each FFI cutout from a TESS sector is accompanied to its right by the de-trended light curve.}
    \label{fig:tess_indiv}
\end{figure}

In Figure \ref{fig:tess_indiv} we show the FFI cutouts and the de-trended light curve for each available sector for TIC 198459831, the same source shown in the ZTF light curve figure (Figure \ref{fig:ztf_summary_plot}). We note that while there is evidence of artifacts/unremoved systematic effects in the de-trended light curves, especially in the regions adjacent to the mid-sector data gap, as we show in the following section it is still possible to recover periodic signals in these light curves.

\subsection{Offset-corrected Periodic Signal Search} \label{sec:ls_search}
After de-trending, we search for periodic signals by using a modified version of the LS periodogram to account for the flux offsets which exist between sectors. Flux offsets can be caused by insufficient/incorrect de-trending but also by observational limitations due to the source falling on different pixels each sector and the pixel sensitivities not being entirely uniform. 

As discussed in Sections 5 \& 6 of \citet{vanderplas2018}, the LS periodogram is identical to fitting a sinusoid to the data at each frequency/period and can thus be interpreted as least-squares method. Assuming that the data has been normalized by its mean/median, the model can be written as $y(t) = w_1 \sin\left(\frac{2\pi}{P}t\right) + w_2 \cos\left(\frac{2\pi}{P}t\right)$ where $y(t)$ is the flux value as a function of time $t$, $P$ is the period, and $w_i$'s are the coefficients we are solving for. For a given value of $P$, this model is linear and the coefficients $w_i$ can be solved for in closed-form. In matrix form, the model can be written as 

\begin{equation}
    \boldsymbol{y} = \boldsymbol{X}_\mathrm{LS}\cdot \boldsymbol{w} + \boldsymbol{\varepsilon}
\end{equation}
where $\boldsymbol{y}, \boldsymbol{\varepsilon}$ are $N \times 1$ column vectors containing the $N$ flux measurements $y_i$ and their associated uncertainties $\varepsilon_i$, and $\boldsymbol{w}$ is a $2 \times 1$ column vector containing $w_1, w_2$. The design matrix $\boldsymbol{X}_\mathrm{LS}$ is of shape $N \times 2$ and can be written as  

\begin{equation}
    \boldsymbol{X}_\mathrm{LS} = 
    \begin{bmatrix}
    \sin\left(\frac{2\pi}{P}t_1\right) & \cos\left(\frac{2\pi}{P}t_1\right) \\ 
    \sin\left(\frac{2\pi}{P}t_2\right) & \cos\left(\frac{2\pi}{P}t_2\right) \\
    \vdots & \vdots \\
    \sin\left(\frac{2\pi}{P}t_N\right) & \cos\left(\frac{2\pi}{P}t_N\right) \\
    \end{bmatrix}.
\end{equation}
The ordinary least-squares solution is 
\begin{equation} \label{eq:ols_sol}
    \boldsymbol{\hat{w}} = (\boldsymbol{X}_\mathrm{LS}^\top \cdot \boldsymbol{X}_\mathrm{LS})^{-1} \cdot (\boldsymbol{X}_\mathrm{LS}^\top \cdot \boldsymbol{y})
\end{equation}
and the model prediction is given by $\boldsymbol{\hat{y}} = \boldsymbol{X}_\mathrm{LS} \cdot \boldsymbol{\hat{w}}$. To calculate the LS power at a given period $P$ with the standard normalization\footnote{\url{https://docs.astropy.org/en/stable/timeseries/lombscargle.html}} used by \texttt{Astropy}, we use the formula

\begin{equation}
    \mathtt{power}(P) = \frac{\chi^2_\mathrm{ref} - \chi^2(P)}{\chi^2_\mathrm{ref}}
\end{equation}
where $\chi^2(P) = \sum_i \left( y_i - \hat{y_i}\right)^2$ and in our case we set $\chi^2_\mathrm{ref} = \sum_i \left(y_i  - \frac{1}{N}\sum y_i\right)^2$. This normalization is also referred to as the ``least-squares" normalization in Section 7.5 of \citealt{vanderplas2018}. By calculating the model and $\chi^2$ for each period we can obtain the power. 

The least-squares interpretation allows us to modify the periodogram by adding elements to the design matrix to model more variations in the data. This approach of augmenting the design matrix to simultaneously fit the signal of interest and systematics was also used in \citet{dfm2015} and \citet{angussip2015} where they used ``eigen light curves" to model systematic effects in Kepler data. In our case, the design matrix can be modified to account for the flux offsets by adding separate rectangle/boxcar functions that have nonzero values for each sector. Fitting for these is \textit{not} the same as normalizing the flux values, which we also perform, but to correct for artificial offsets created at the boundaries between light curves due to individually normalizing them. While these offsets would not exist for a relatively quiet star, for variable stars it is possible that these artifacts suppress a periodic signal longer than an observational sector. 
Using the notation $T_{m,0}, T_{m,f}$ to be the time the observations started and ended for the $m$-th sector, we can write the rectangle function for the $m$-th sector as  

\begin{equation}
    \Pi_m(t) = 
    \begin{cases}
       1, & T_{m,0} \leq t \leq T_{m,f}, \\ 
       0, & \mathrm{otherwise}\\
    \end{cases}
\end{equation}
which is a $N \times 1$ column vector (where $N$ is the total number of observations across multiple sectors) with nonzero values only during the time that the $m$-th sector is observed. 
We can write the collection of $M$ step functions in matrix notation as 

\begin{equation}
    \boldsymbol{\Pi} = 
    \begin{bmatrix}
        \Pi_1(t_1) & \Pi_2(t_1) & \cdots & \Pi_M(t_1) \\
        \Pi_1(t_2) & \Pi_2(t_2) & \cdots & \Pi_M(t_2) \\
        \vdots & \vdots & \vdots & \vdots \\ 
        \Pi_1(t_N) & \Pi_2(t_N) & \cdots & \Pi_M(t_N) \\
    \end{bmatrix}.
\end{equation}

We can combine the LS design matrix and the rectangle function matrix to create a new design matrix
\begin{equation}
    \boldsymbol{X} = 
    \begin{bmatrix}
        \boldsymbol{X}_\mathrm{LS} & \boldsymbol{\Pi}
    \end{bmatrix}
\end{equation}
of shape $N \times (2 + M)$ containing features to model a sinusoidal signal in the presence of flux offsets between sectors. As this model is still linear for a given value of $P$, it permits the same closed-form formula as Equation \ref{eq:ols_sol} (now with $\boldsymbol{X}$ instead of $\boldsymbol{X}_\mathrm{LS}$). Fitting for a sinusoidal signal and the flux offsets simultaneously allows us to prevent artifacts and errors which can be introduced when performing these steps separately. 

\begin{figure}
    \centering
    \includegraphics[width=\textwidth]{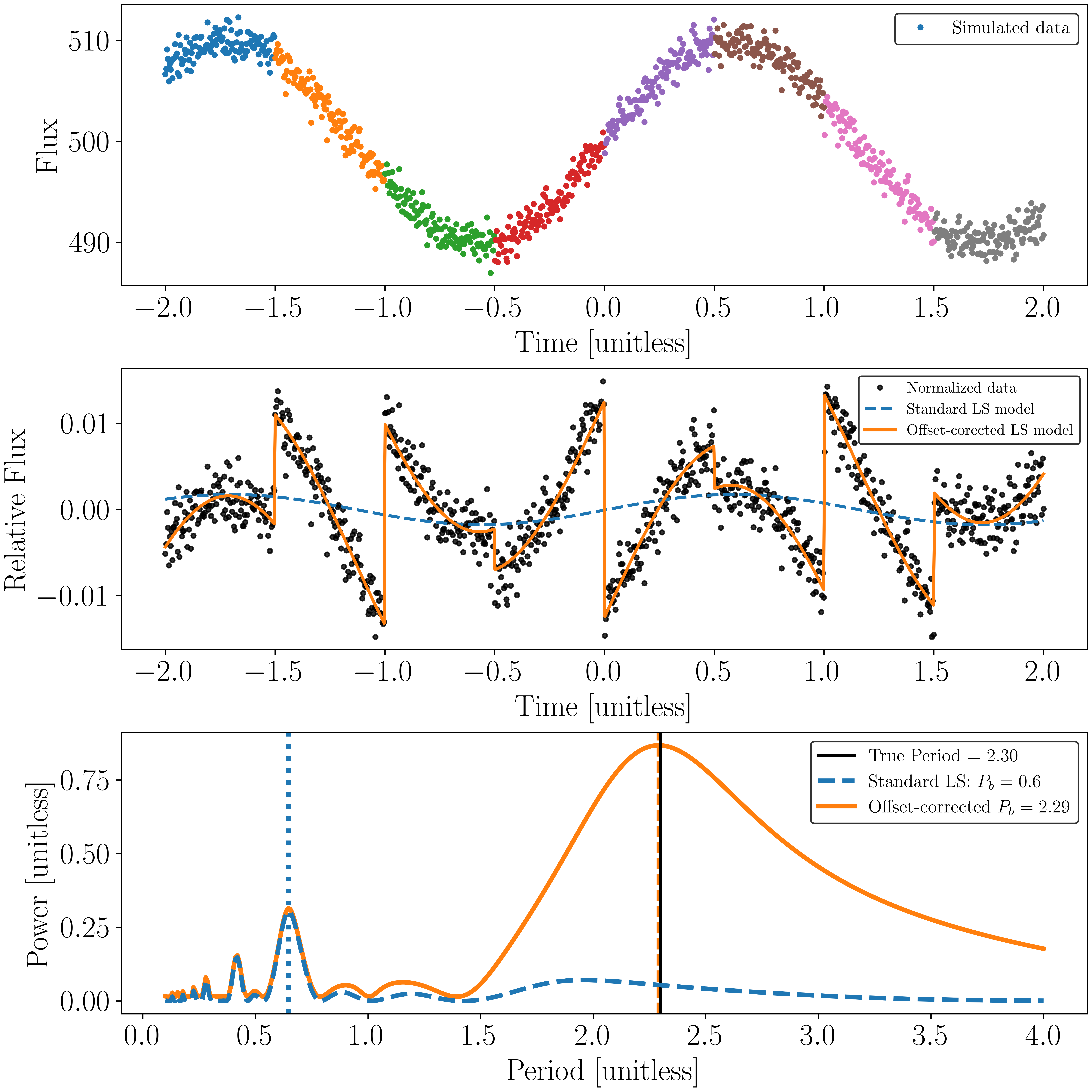}
    \caption{A toy example showing the offset-corrected LS periodogram. The top panel shows a simulated sinusoidal signals with some Gaussian noise. The simulated light curve is broken up into eight equal chunks, represented by different colors in the top panel, similar to how a TESS target might be observed each sector. In the middle panel we show in black what the light curve would look like if each chunk was normalized according to its own median value. We subtract 1 from each light curve chunk to center it at zero. The dotted blue line shows the best-fit model obtained by running a standard LS periodogram. The orange solid line shows the best-fit model using the offset-corrected LS periodogram. In the bottom panel we show the obtained periodogram from each of the models (i.e., standard vs offset-corrected). See Section \ref{sec:ls_search} for more information on the offset-corrected LS.}
    \label{fig:offset_corrected_LS}
\end{figure}

We show a simulated toy example of this offset-corrected LS periodogram in Figure \ref{fig:offset_corrected_LS}. We generated a sinusoidal signal with Gaussian noise and split it up into eight chunks (top panel). While we do not add flux offsets between each chunk, in practice those would also likely be present in the data.
We then normalized and centered each chunk of the light curve separately, similar to how one might do so for multiple TESS observations (middle panel). We can see that normalizing the chunks separately leads to artifacts at the boundaries. While the standard LS periodogram (blue dashed line in bottom panel) is unable to detect the true period, the offset-corrected LS (orange solid line) is able to recover the true period of the sinusoid. While this example is extreme, it nonetheless showcases how potential artifacts can suppress sinusoidal signals in the data and how this simple modification can alleviate the problem. While a natural extension of this work would be to create a design matrix to incorporate the \texttt{unpopular} systematics correction, flux offset correction, and the sinusoidal signal search, we leave this approach for a future study.

\begin{figure}
    \centering
    \includegraphics[width=\textwidth]{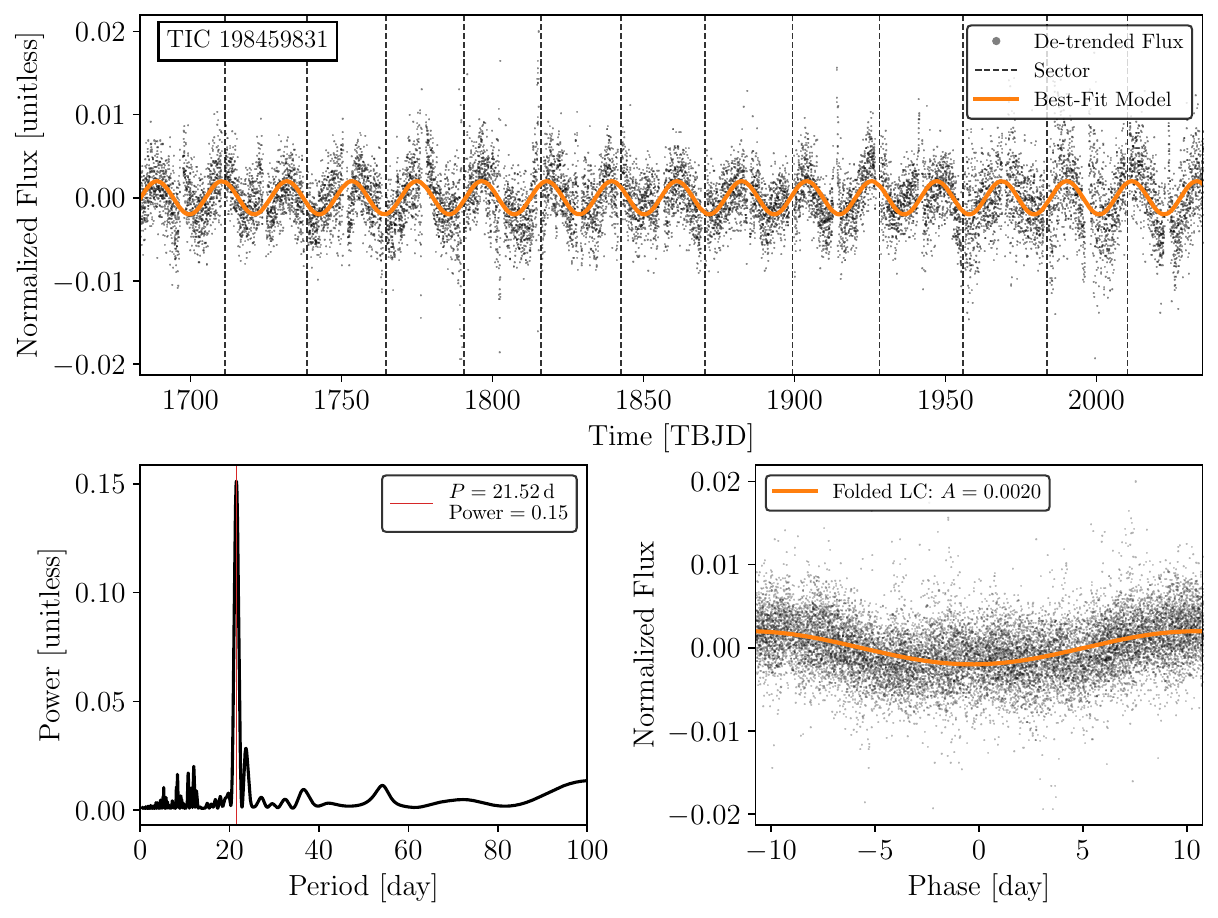}
    \caption{The de-trended TESS light curve and period analysis for TIC 198459831. In the top panel we show the de-trended normalized and centered de-trended flux measurements (black dots), the sector boundaries(dashed lines) , and the best-fit model (orange line). In the bottom left panel we show the periodogram and indicate the period of $P=21.52\,\mathrm{day}$ corresponding to the highest power (red line). In the bottom right panel we show the phase folded light curve and the amplitude of the oscillation. We can see that the recovered TESS period matches that measured from the ZTF light curve shown in Figure \ref{fig:ztf_summary_plot}.}
    \label{fig:tess_summary}
\end{figure}

As mentioned above, while we median normalize and center each sector's flux value to remove the bulk of the offsets, incorporating these flux offsets directly into our model allows us to more accurately capture a sinusoidal signal in the presence of residual flux offsets. In Figure \ref{fig:tess_summary} we show our de-trended TESS light curve and periodic signal search. TIC 320504531 (Gaia ID 1434665625444326016) has a Gaia mean $G$ magnitude of 14.84, a TESS magnitude of 13.6, and an estimated $T_\mathrm{eff}$ of $3342\pm157$ K.
for TIC 198459831, the same source presented above for the ZTF light curve (see Figure \ref{fig:ztf_summary_plot}). TIC 198459831 (Gaia ID 1434755678023105408) has a Gaia mean $G$ magnitude of 14.50, a TESS magnitude of 13.37, and an estimated $T_\mathrm{eff}$ of $3509\pm157$ K.
We evaluate the offset-corrected LS periodogram over 5000 equally spaced periods between 0.5-100 days. We included search periods that are shorter than what we may except given ZTF's measured period to prevent biasing our TESS-measured periods to being longer than 10 days. The recovered period from the TESS light curve is 21.52 days, agreeing with the ZTF-derived period of 21.51 days. To have a quantitative grasp on how reliable this approach is, we repeated this periodic search for all 271 remaining sources in the sample.

\section{Comparison of Periods} \label{sec:results}

\begin{figure}
    \centering
    \includegraphics[width=\textwidth]{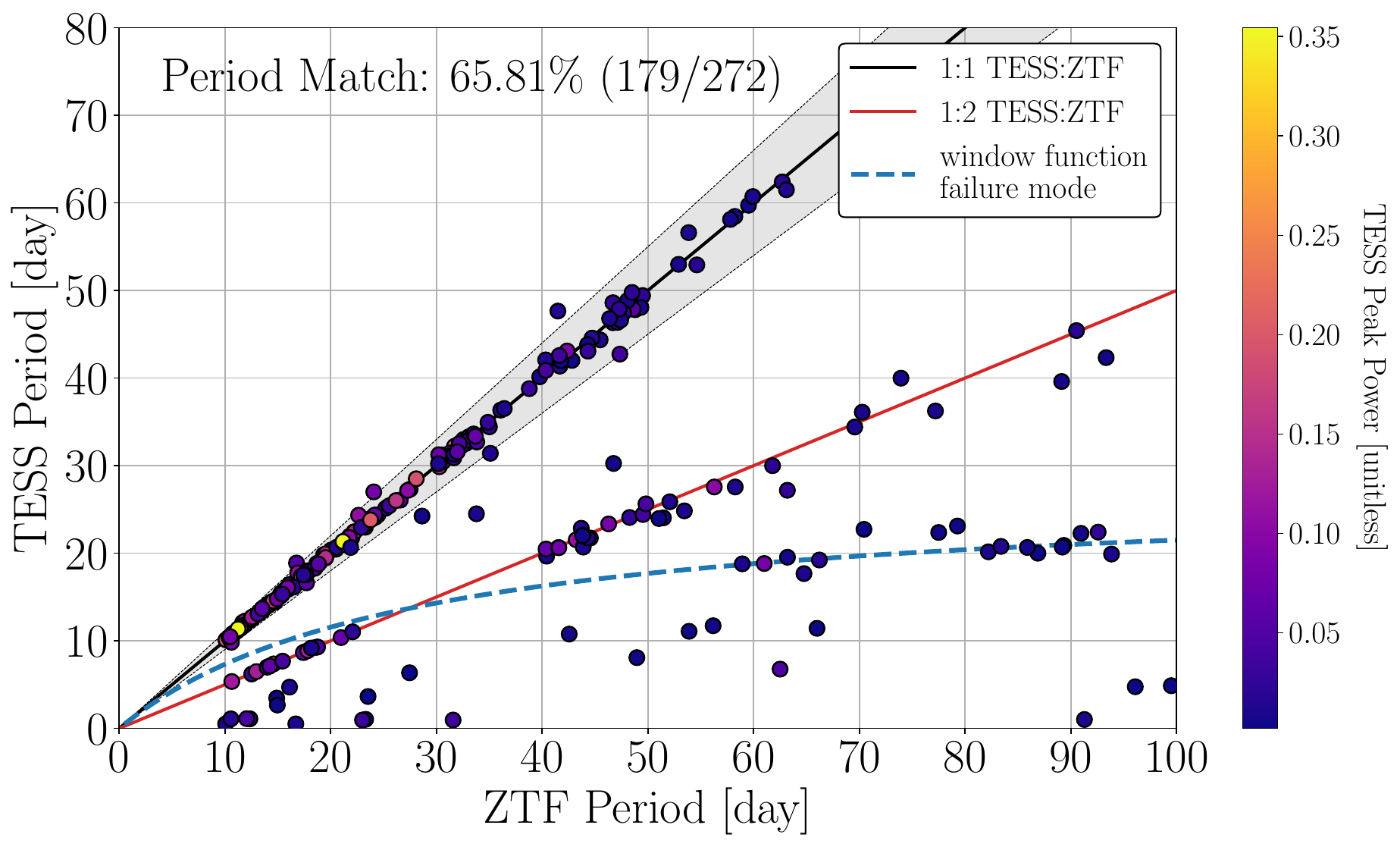}
    \caption{A plot comparing the periods measured from the ZTF light curves and the TESS light curves. The points are colored by their TESS peak powers. The 1:1 line with 10\% uncertainties are indicated with the black line and the shaded region. The 1:2 line is indicated in red. A potential Lomb-Scargle failure mode caused by the window function of the TESS observing sector is indicated as a dashed blue line (more description in text).}
    \label{fig:ztf_tess_comparison}
\end{figure}

In this section we show our results of how well we can recover the same periods from TESS and the ZTF light curves for the 272 sources in our sample. In Figure \ref{fig:ztf_tess_comparison} we show the ZTF \& TESS measured periods for each source in our sample. A pair of ZTF \& TESS periods are considered to be matching if the fractional difference between the periods is less than 10\% (i.e., $\frac{\vert P_\mathrm{ZTF} - P_\mathrm{TESS}\vert}{P_\mathrm{ZTF}} < 0.1$). Of the 272 sources, 179 (66\%) of them having matching periods and lie along the 1:1 line (black line).  Of the 93 mismatches, 43 sources (16\% of the 272 sources) lie on the 1:2 TESS:ZTF double harmonic line in red (i.e., the TESS periodogram attains a max power at double the frequency/half the period of that of ZTF). For the other mismatches, we also briefly note that several of them lie close to a LS periodogram failure mode (blue dashed line) that is caused by the interaction of the window function of a TESS observing sector with $P_\mathrm{sector}=27.4$ days and a given source's period (see Section 7.2 of \citealt{vanderplas2018}).

\begin{figure}
    \centering
    \includegraphics[width=0.8\textwidth]{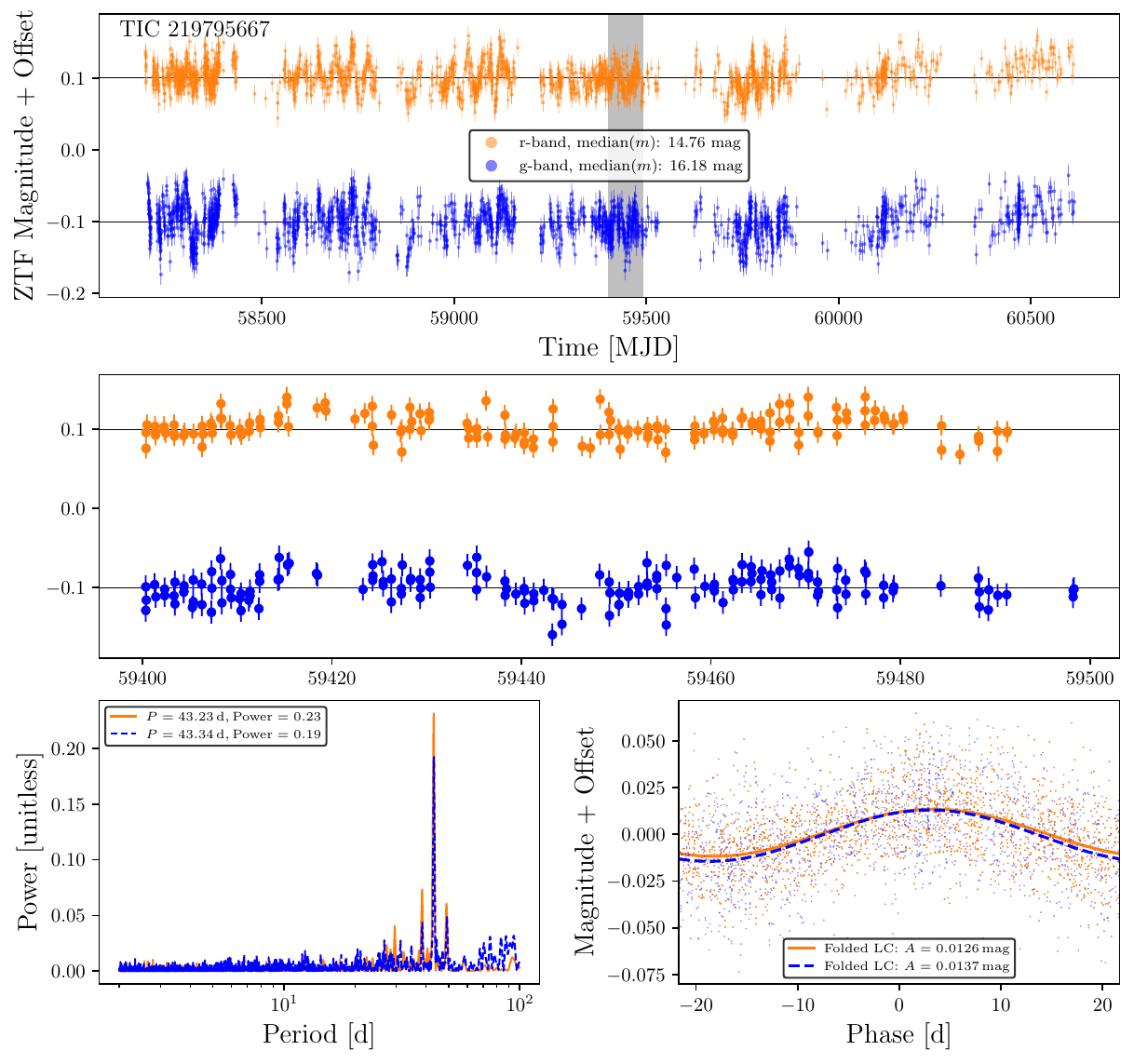}
    \includegraphics[width=0.8\textwidth]{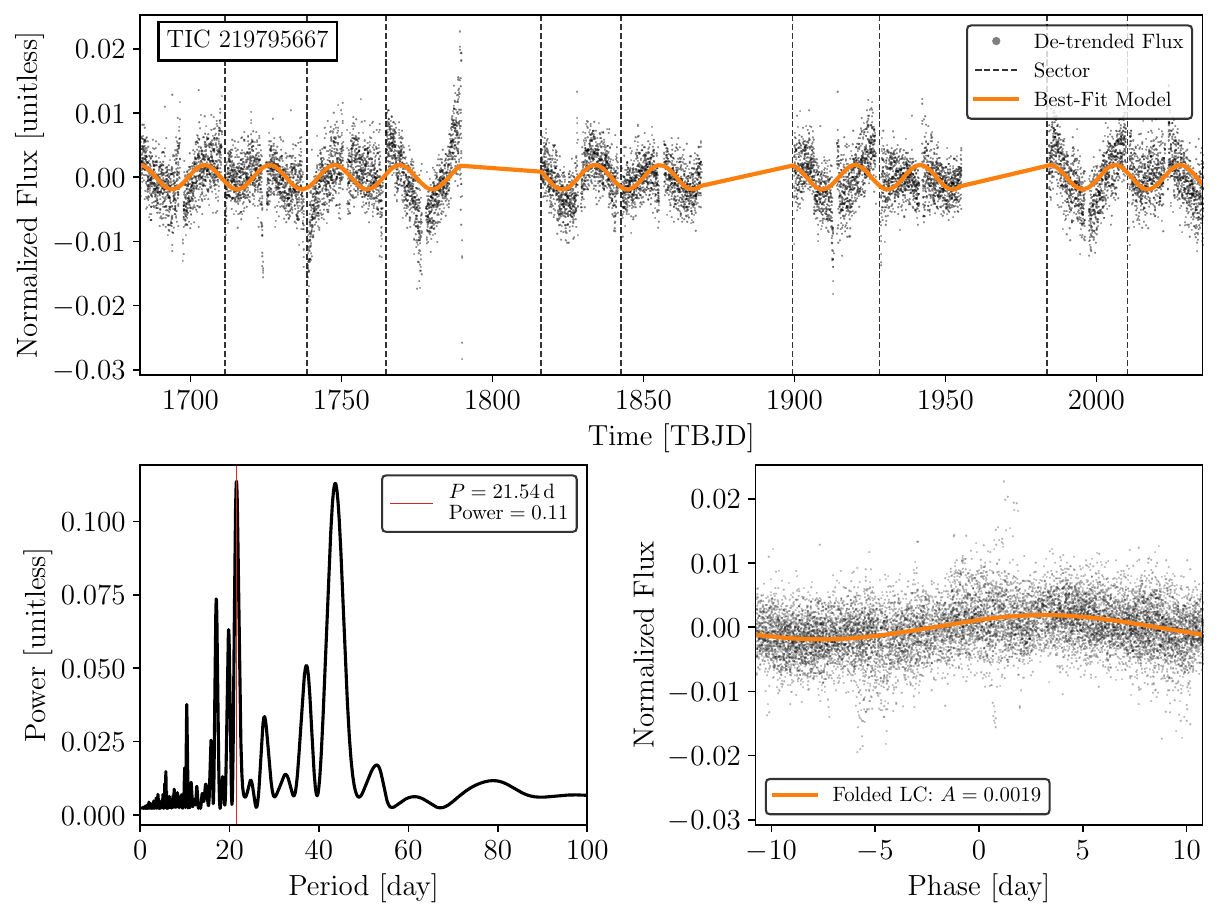}
    \caption{A source (TIC 219795667) where we measured a TESS period that is half of what was measured from ZTF. We can see that the second TESS periodogram peak at 43 days corresponds to the same period as that measured from ZTF.}
    \label{fig:one_to_two}
\end{figure}

There are at least two scenarios that cause the 43 sources to exist on the double harmonic line. The first scenario is caused by a common failure mode when LS periodograms are applied to periodic signals that are not strictly sinusoidal; brightness variations caused by stellar rotation are rarely strictly sinusoidal. For these non-strictly sinusoidal signals, even if the fundamental frequency is $f_0$ there tends to be power at higher harmonics such as $2f_0$. The LS periodogram can then erroneously calculate a larger power at the higher harmonic instead of the true frequency (see Section 7.2 of \citealt{vanderplas2018}).
We show an example of such a source, TIC 219795667, in Figure \ref{fig:one_to_two} where we measure a period of ${\sim}$43 days with the ZTF light curves but a period of $\sim$22 days with the TESS light curve. Looking at the TESS periodogram, we can see that at $P=43$ days we also a peak in the power, which would match the period measured by ZTF. This discrepancy is likely a combination of the incomplete systematics correction in the TESS light curve, as evidenced by variations that repeat at a sector length, and of measuring a higher order harmonic for non-strictly sinusoidal signals. Given how close the TESS LS powers are for the two periods, it is possible that de-trending the TESS light curves with a different set of hyperparameters may have led to a higher peak power at the ZTF-measured period of $P=43$ days. We plan to explore varying the hyperparameters in a future study.

\begin{figure}
    \centering
    \includegraphics[width=0.8\textwidth]{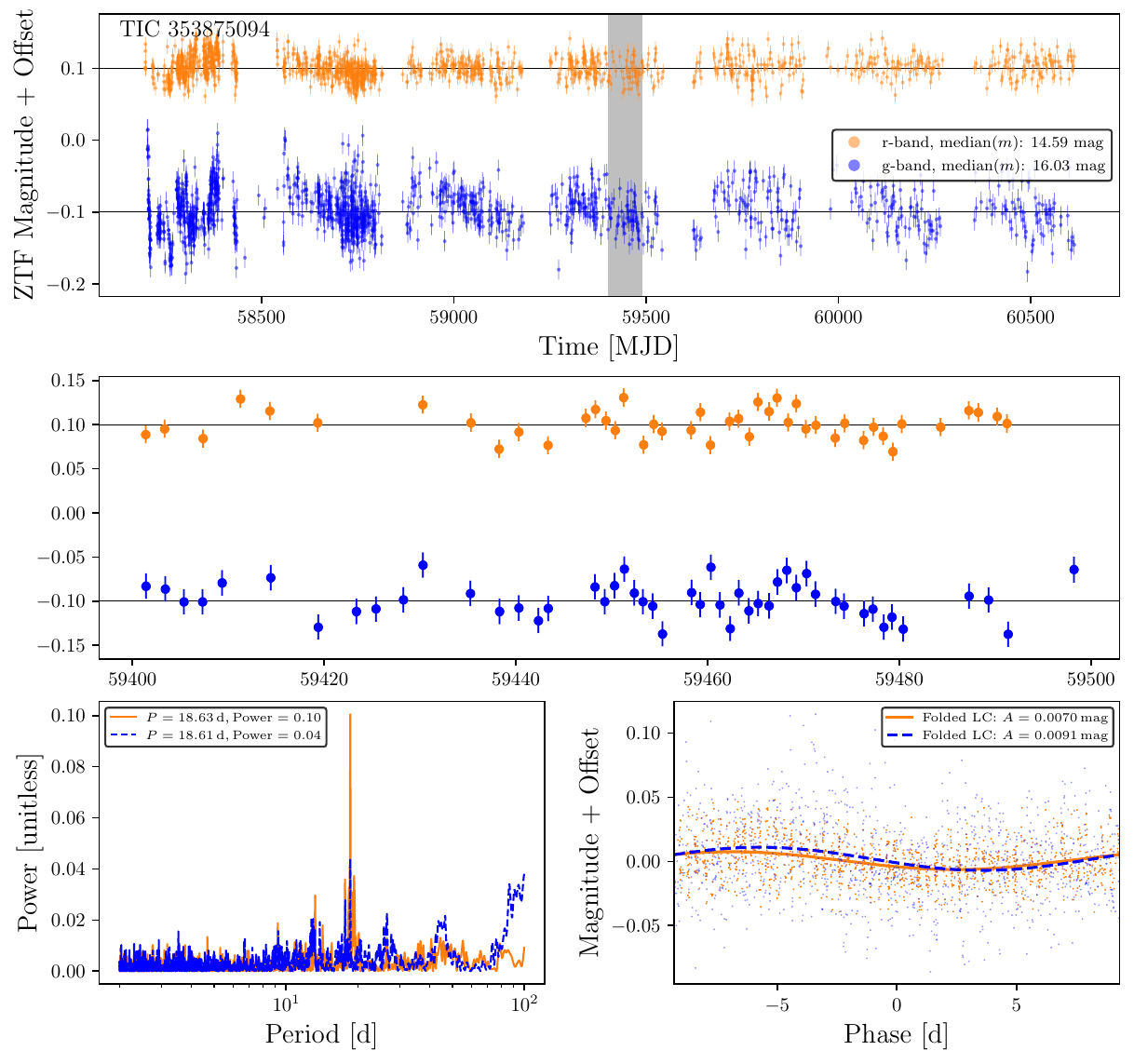}
    \includegraphics[width=0.8\textwidth]{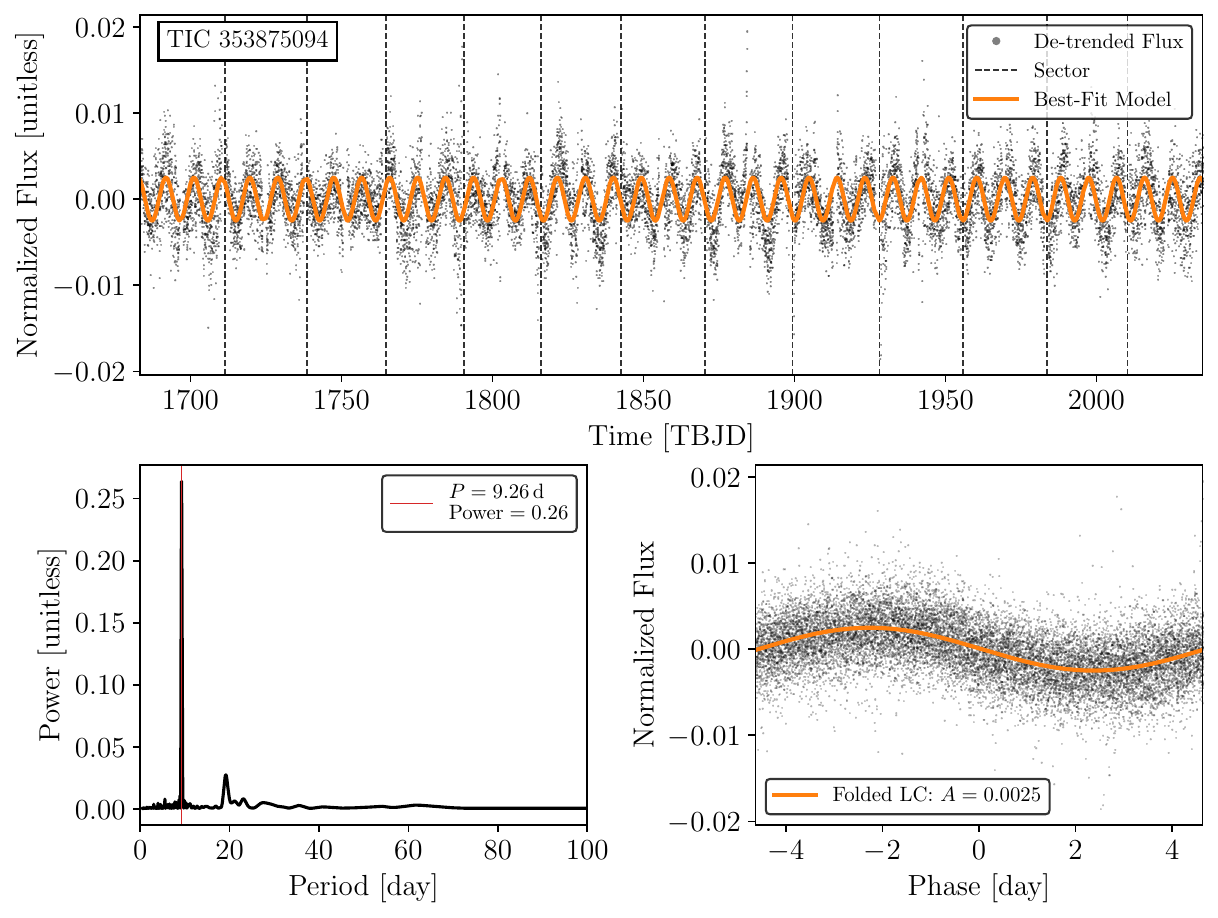}
    \caption{A source (TIC 353875094) where the shorter TESS-measured period of $P_{\mathrm{TESS}}=9.26$ days is likely the correct rotation period instead of the ZTF-measured period of $P_{\mathrm{ZTF}}=18.63$ days. In the TESS periodogram we see that while there is a small peak at $P_\mathrm{ZTF}$, the peak at $P_\mathrm{TESS}$ is much stronger.  
    }
    \label{fig:shorter_tess_correct}
\end{figure}

The other scenario, likely for the sources that the TESS-measured period is shorter than 10 days, is that ZTF erroneously measured a longer period. It is difficult to identify the conditions under which ZTF periods are erroneously wrong, but a possible explanation is that for shorter period signals TESS's higher precision and 30-minute cadence leads to more accurate period measurements. We show an example of this scenario in Figure \ref{fig:shorter_tess_correct}, where the TESS-measured period of $P_{\mathrm{TESS}}=9.26$ days is likely the correct rotation period instead of the ZTF-measured period of $P_{\mathrm{ZTF}}=18.63$ days. Of the 43 sources on the double harmonic line, for 13 sources TESS measures a period shorter than 10 days.

If we were to also tentatively consider these 1:2 cases as matches, as the periodograms are picking up an alias of some ``true" period, we would have  an upper-limit match rate of 82\% (222/272).

\begin{figure}
    \centering
    \includegraphics[width=\textwidth]{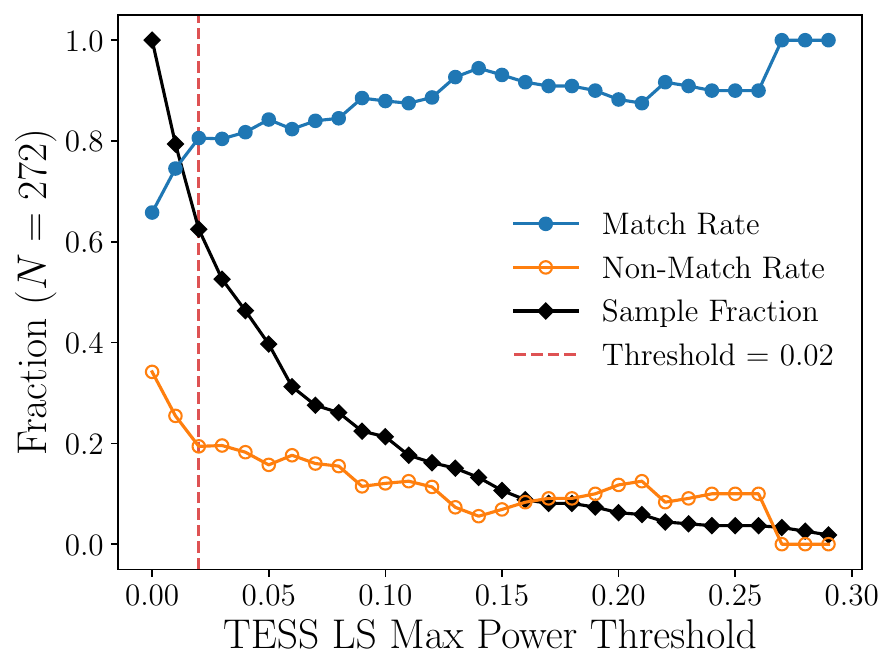}
    \caption{A plot showing the match (blue)and non-match (orange) rates as a function of the TESS power threshold. The black line indicates the fraction of remaining samples as we increase the power threshold.}
    \label{fig:power_threshold}
\end{figure}

\begin{figure}
    \centering
    \includegraphics[width=\textwidth]{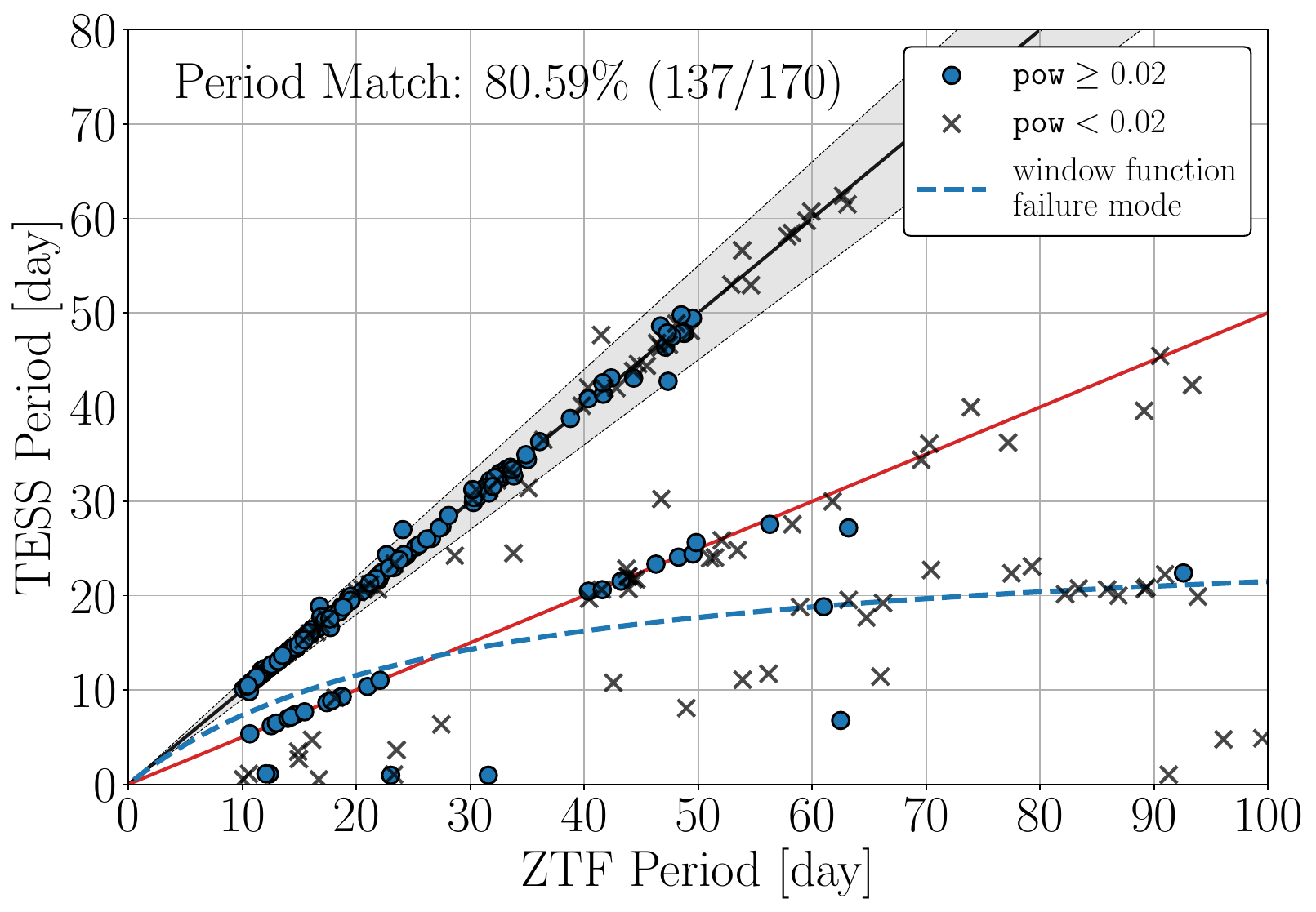}
    \caption{Same figure as Figure \ref{fig:ztf_tess_comparison} but with the power cut of \texttt{power} $>0.02$ applied. While this removes a significant fraction of the samples, we see that it removes many of the mismatched periods in the lower right of Figure \ref{fig:ztf_tess_comparison}, increasing our match percentage.}
    \label{fig:ztf_tess_comparison_cut}
\end{figure}

We then explored performing a cut based on the TESS LS max power at the measured period to increase the percentage of ZTF-TESS period matches. At each power threshold, where we started from 0 and increased in increments of 0.01 until we reached 0.3, we removed the sources where the TESS LS max power was smaller than the threshold value from the  sample. In Figure \ref{fig:power_threshold} we show the results of this experiment. The objective we are pursuing with this power threshold is to increase the match rate and decrease the non-match rate without removing too many sources from our sample. Therefore, we would like to use the smallest power threshold while increasing the match-rate. This TESS power threshold turns out to be $\texttt{power=0.02}$, where out of the 170 sources left in our sample 137 of them having matching periods (81\%) (see Figure \ref{fig:ztf_tess_comparison_cut}). If we were to count the remaining 1:2 harmonics ($N=22$) as matches like above, 94\% of our sources have matching TESS-ZTF periods. While the cut removed some sources with matching periods between 40 \& 50 days, it also removed many of the sources with mismatched periods. This result shows that for studies that involve creating a large catalog of potential long rotation periods from TESS light curves for which the LS powers are also reported, the catalog user can perform a fidelity cut based on the requirements of their study. 

Table \ref{tab:match_fractions} shows the fraction of sources that lie on the TESS:ZTF 1:1 line, 1:2 line, and neither of those lines for both the subset of the sample that passed the power threshold (i.e., high TESS power sample) described in the previous paragraph and the subset that did not (i.e., low TESS power sample). Of the 43 sources that lie on the TESS:ZTF 1:2 line half of the sources are still in the high TESS power sample, while for the remaining 50 mismatches only 11 sources remain in the high TESS power sample. This result indicates that the thresholding has a stronger effect in identifying and removing sources where the systematics in the TESS light curves were not adequately removed relative to its effect in removing sources with higher-order harmonics. Varying the hyperparameters in the \texttt{unpopular} TESS de-trending pipeline and understanding how it affects the maximum TESS LS power is a potential future direction for this project.

\begin{table}[]
\begin{centering}
    
\begin{tabular}{|c|c|c|c|}
\hline
                          & TESS:ZTF 1:1      & 1:2              & Neither          \\ \hline
High TESS Power $(N=170)$ & $81\%\,(137/170)$ & $13\%\,(22/170)$ & $6\%\,(11/170)$  \\ \hline
Low TESS Power $(N=102)$  & $41\%\,(42/102)$  & $21\%\,(21/102)$ & $38\%\,(39/102)$ \\ \hline
Total Sample $(N=272)$    & $66\%\,(179/272)$ & $16\%\,(43/272)$ & $18\%\,(50/272)$ \\ \hline
\end{tabular}
    \caption{Table showing the fraction of TESS:ZTF 1:1 matches, 1:2 matches, and those that are neither for the subset of the sample that passes the TESS power threshold cut (the high TESS power sample), the subset of the sample that does not pass the cut (the low TESS power sample), and the total sample.}
    \label{tab:match_fractions}
\end{centering}

\end{table}

\begin{figure}
    \centering
    \includegraphics[width=0.8\textwidth]{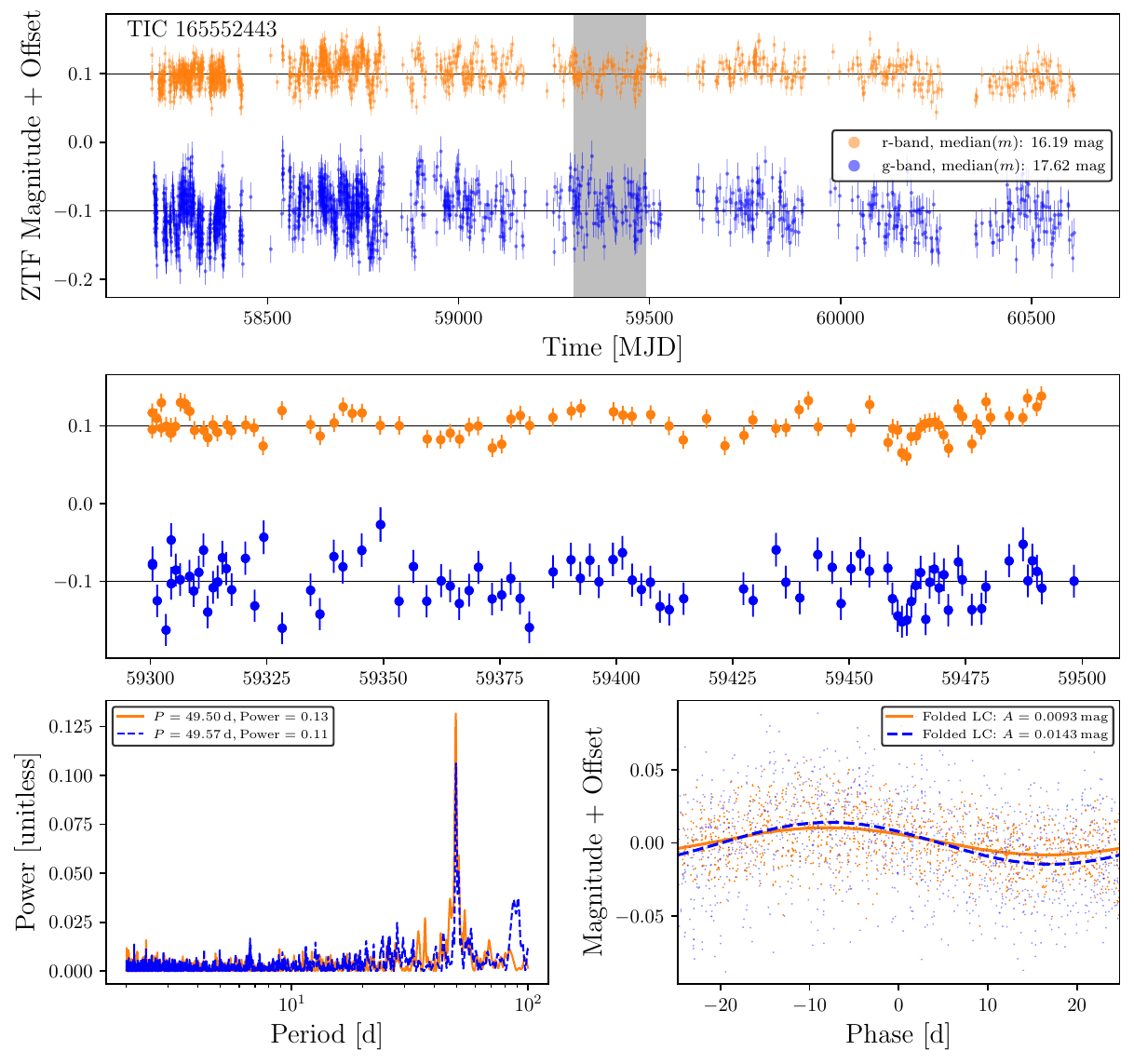}
    \includegraphics[width=0.8\textwidth]{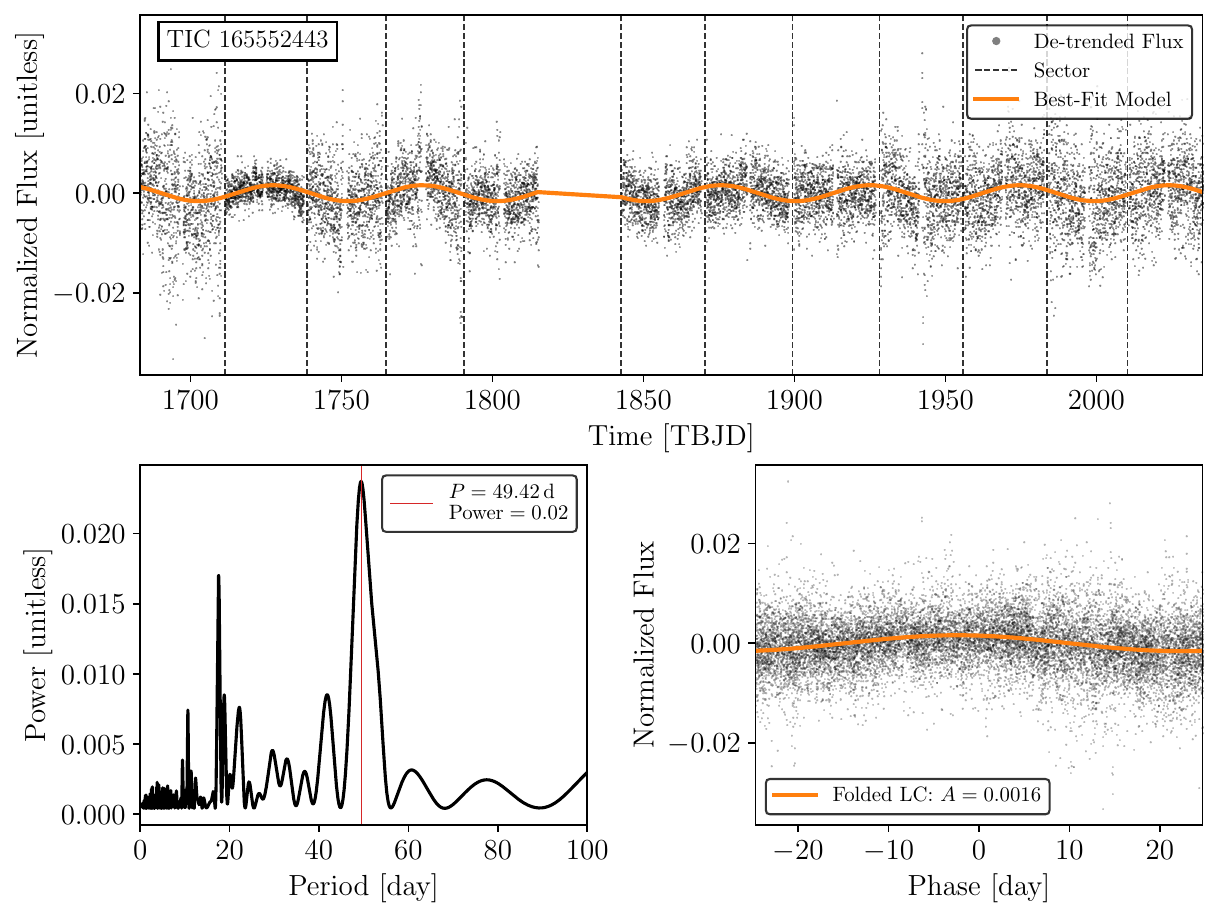}
    \caption{An example of a source (TIC 165552443) where we measured a period of 50 days from both ZTF and TESS light curves. As an individual TESS sector is 27 days, this result shows it is possible to probe for periodic signals exceeding the length of a single sector.}
    \label{fig:long_period}
\end{figure}

While Figure \ref{fig:ztf_tess_comparison_cut} shows that most of the period matches occur between 10 \& 30 days, there are several sources with longer periods that lie on the 1:1 line. As an example, in Figure \ref{fig:long_period} we show a case where we recovered a matching period of ${\sim}50$ days from ZTF and TESS where the TESS peak power would pass the threshold power of 0.02. 

\section{Conclusion \& Future Work} \label{sec:conclusion}
We have presented results showing that it is possible to measure periodic signals beyond 10 days using classical periodogram approaches from TESS prime-mission FFI light curves for sources in the NCVZ and even beyond the length of a single TESS sector (27 days). We developed a sample from the TESS Candidate Target List which passed the criteria presented in Section \ref{sec:ztf}. We then downloaded the $r$ \& $g$ band (when available) light curves from the ground-based long (6+ year) baseline observations of ZTF. We performed a periodic signal search on these targets and retained the targets which had reliable ZTF-measured periods longer than 10 days of which there were $N=272$ sources. In the absence of a large catalog of long-period rotators, we consider the ZTF periods to be the ``ground truth" periods, given their long observational baseline, that we would like to recover using TESS light curves. 
For each of these sources we then downloaded the available TESS Cycle 2 FFI cutouts from the prime mission and de-trended each sector using the \texttt{unpopular} pipeline. To account for flux offsets between sectors, we modified the LS periodogram to simultaneously fit for these offsets while searching for a periodic signal (Section \ref{sec:ls_search}). We showed that we can measure the same period from the TESS light curves as those from ZTF for 179 out of 272 sources (66\%). By doing a power cut based on the TESS LS power and removing sources that do not pass this threshold from the sample, we showed that 137 out of 170 sources (81\%) have matching periods between 10-50 days. 

One major limitation of this study is that we chose a sample with relatively strong sinusoidal signals based on the ZTF light curves. While this selection was adequate for this study as our objective was to provide evidence that periodic signals longer than 10 days can be recovered from TESS light curves as opposed to a new catalog of rotation periods measured from TESS, we caution with the interpretation of the match rates provided here. While it is tempting to interpret them as the fraction of sources with reliable periods measured using TESS FFI light curves, a more accurate interpretation is \textit{the fraction of sources where we can measure reliable periods when they exhibit relatively strong periodic signals}. Another limitation of this study is that while we would like to measure stellar rotation periods as opposed to any sinusoidal signal, it is difficult to make that distinction with just this simple analysis. As stellar oscillations tend to be quasi-periodic, a more accurate approach would be to replace the sinusoidal signal search with a Gaussian process with an appropriate kernel. 

While many rotation period studies using TESS have focused on periods shorter than 10 days due to the spacecraft's unique systematics, these results show that it is worth exploring longer periods. Probing longer periods with TESS would dramatically increase our sample of long-period rotators, potentially leading to new discoveries of stellar evolution and also low-mass dwarfs. A future study will provide a catalog of potential long-period rotators from the TESS CVZs using this approach.  

\newpage
\section*{Acknowledgments}
We would like to thank the anonymous referee for constructive feedback that improved the paper.
This paper includes data collected by the TESS mission. Funding for the TESS mission is provided by the NASA's Science Mission Directorate. All the TESS data used in this paper can be found in MAST: \dataset[10.17909/fwdt-2x66]{https://doi.org/10.17909/fwdt-2x66} \citep{TESS_data}.
This work has made use of data from the European Space Agency (ESA) mission {\it Gaia} (\url{https://www.cosmos.esa.int/gaia}), processed by the {\it Gaia} Data Processing and Analysis Consortium (DPAC, \url{https://www.cosmos.esa.int/web/gaia/dpac/consortium}). Funding for the DPAC has been provided by national institutions, in particular the institutions participating in the {\it Gaia} Multilateral Agreement. This work has made use of data based on observations obtained with the Samuel Oschin Telescope 48-inch and the 60-inch Telescope at the Palomar Observatory as part of the Zwicky Transient Facility project. ZTF is supported by the National Science Foundation under Grants No. AST-1440341 and AST-2034437 and a collaboration including current partners Caltech, IPAC, the Oskar Klein Center at Stockholm University, the University of Maryland, University of California, Berkeley , the University of Wisconsin at Milwaukee, University of Warwick, Ruhr University, Cornell University, Northwestern University and Drexel University. Operations are conducted by COO, IPAC, and UW. The ZTF data used in this paper can be found in IPAC/IRSA: \dataset[10.26131/IRSA539]{https://doi.org/10.26131/IRSA539} \citep{ZTF_data}.
RA acknowledges support from NASA grant 80NSSC21K0636.

%

\vspace{5mm}
\facilities{ZTF, TESS, Gaia}


\software{\texttt{AstroPy} \citep{astropy:2018}, \texttt{lightkurve} package \citep{barentsen2020}, \texttt{TESScut} \citep{brasseur2019},
\texttt{NumPy} \citep{harris2020array}, 
\texttt{Matplotlib} \citep{Hunter:2007}
          }





\newpage
\bibliography{ms}{}
\bibliographystyle{aasjournal}



\end{document}